\documentclass[apj,iop, graphicx]{emulateapj}
\usepackage{apjfonts}

\DeclareTextFontCommand{\textmyfont}{\myfont}

\makeatletter
\newcommand*{\rom}[1]{\expandafter\@slowromancap\romannumeral #1@}

\newcommand*{\myfont}{\fontfamily{cmtt}\selectfont}

\begin{document}

\title{A Systematic Spectral-Timing Analysis of Kilohertz  Quasi--Periodic Oscillations in the Rossi X-ray Timing Explorer Archive}
\shortauthors{Troyer et al.}
\shorttitle{Spectral--timing analysis of kHz QPOs}

\author{Jon~S.~Troyer\altaffilmark{1}, Edward~M.~Cackett\altaffilmark{1}, Philippe ~Peille\altaffilmark{2},  Didier ~Barret\altaffilmark{3,4}}

\email{jon.troyer@wayne.edu}

\affil{\altaffilmark{1}Department of Physics \& Astronomy, Wayne State University, 666 W. Hancock St, Detroit, MI 48201, USA}
\affil{\altaffilmark{2}Centre National d'\'Etudes Spatiales, Centre spatial de Toulouse, 18 avenue Edouard Belin, 31401 Toulouse cedex 9, France}
\affil{\altaffilmark{3}Universit\'e de Toulouse; UPS-OMP; IRAP; Toulouse, France}
\affil{\altaffilmark{4}CNRS; Institut de Recherche en Astrophysique et Plan\'etologie; 9 Av, colonel Roche, BP 44346, F-31028 Toulouse cedex 4, France}

\begin{abstract}
Kilohertz quasi-periodic oscillations or kHz QPOs occur on the orbital timescale of the inner accretion flow and may carry signatures of the physics of strong gravity (c$^{2}\sim$ GM/R) and possibly clues to constraining the neutron star equation of state (EOS).  Both the timing behavior of kHz QPOs and the time-averaged spectra of these systems have been studied extensively, yet no model completely describes all the properties of kHz QPOs.  Here, we present a systematic study of spectral-timing products of kHz QPOs from low-mass X-ray binary systems using archival Rossi X-ray Timing Explorer/Proportional Counter Array data.  For the lower kHz QPOs in fourteen objects and the upper kHz QPOs in six, we were able to obtain correlated time-lags as a function of QPO frequency and energy, as well as energy-dependent covariance spectra and intrinsic coherence.  For the lower kHz QPOs, we find a monotonic decrease in lags with increasing energy, rising covariance to $\sim$ 12 keV, and near unity coherence at all energies.  For the upper kHz QPOs, we find near zero lags, rising covariance to $\sim$ 12 keV, and less well-constrained coherence at all energies.  These results suggest that while kHz QPOs are likely produced by similar mechanisms across the population of LMXBs, the lower kHz QPOs are likely produced by a different mechanism than upper kHz QPOs.
\end{abstract}
\keywords{accretion, accretion disks --- stars: neutron --- X-rays: binaries}

\section{Introduction}
Kilohertz quasi-periodic oscillations or kHz QPOs \citep[see e.g.,][]{vanderKlis_98} are the highest-frequency oscillations observed in the X-ray power spectrum of neutron star low-mass X-ray binary systems (LMXBs), and were first detected in 1995 after the launch of NASA's {\it Rossi X-ray Timing Explorer} (RXTE) \citep{Bradt_93}.  The presence of two simultaneous kHz QPOs is common in many neutron star LMXBs, and led to the classification of kHz QPOs as either lower or upper depending on their frequency \citep[see e.g.,][]{vanderKlis_97}.  The short timescale of kHz QPOs is consistent with the dynamical timescales of the inner accretion flow around neutron stars ($\sim100 \mu s$).  Thus, the physical origins of kHz QPOs are of great interest because they offer a possible window to the study of strong gravity ($GM/R \sim c^{2}$) and constraints on the physics of neutron stars \citep[see e.g.,][]{Psaltis_08}. 

Timing analysis of neutron star LMXBs has mostly been phenomenological, characterizing QPO properties through fitting Lorentzians, and observing how those properties change with source state.  Many phenomenological correlations have been for kHz QPOs.  These include various dependences of the properties of all types of QPOs \citep[see e.g.,][]{Mendez_99,vanStraaten_00,DiSalvo_01,DiSalvo_03, Altamirano_08} on the position of the source on a color-color diagram \citep[see e.g.][]{Hasinger_89}, which describes the state of the source.  Any model that we hope to apply to kHz QPOs must explain in a global sense, the many correlations between kHz QPO frequency and other lower frequency oscillations and also the correlations between source state and kHz QPO appearance and properties.  While many models of kHz QPOs have been developed over the years since their discovery \citep[see][for a review]{vanderKlis_00, vanderKlis_06}, none satisfactorily explains all the properties of kHz QPOs.    

Studies involving analyzing the spectral and timing properties of kHz QPOs together can provide a fuller picture of their properties.  However, until relatively recently, timing and spectral analysis approaches have largely been separate, and spectral analysis of neutron star LMXBs has usually focused on fitting physical models to the time-averaged energy spectrum, including the disk, boundary layer, corona etc.  Performing spectral-timing analyses allows one to connect which components of the energy spectrum are related to the QPO, and how those components are connected to each other.  For instance, this can be done through studying the energy-dependence of the variability with a root mean square (rms) spectrum (or similarly a covariance spectrum).  Moreover, time lags between the variability at different energies provide further constraints on the emission processes involved.  Both the covariance spectrum and time lags can be isolated for just the Fourier frequencies covered by the kHz QPOs \citep[see][for a review of spectral-timing techniques]{Uttley_14}.

\begin{deluxetable*}{ccccc}
\tablewidth{0pt}
\tablecolumns{5}
\tablecaption{Neutron Star Low-Mass X-ray Binaries Considered for Analysis}
\tablecomments{Source Classification from \citet{Liu_07}.  While all objects considered are present in the {\it RXTE/PCA} archive, additional columns indicate the presence of observation files in the Event mode necessary for analysis.  Additionally, where event mode files are present, we indicate when these have the full {\it RXTE/PCA} energy range available.  Finally, we also show whether or not we were able to detect kHz QPOs from these observations in the manner described in Section~\ref{data_reduction}.  In some cases, while we were able to detect kHz QPOs, there was not sufficient detected QPO observation time to obtain spectral--timing results.  Those sources shown in bold are the ones where we present analysis.}
\tablehead{
\colhead{Object ID} & \colhead{Source Classification} & \colhead{Event Mode Obs. Present?} & \colhead{Full Energy Range Available?} & \colhead{kHz QPOs Detected?$^{1}$}}
\startdata
{\bf 4U~1608$-$52} & Atoll & Yes & Yes & Yes\\ 
{\bf 4U~1636$-$53} & Atoll & Yes & Yes & Yes\\ 
{\bf 4U~0614$+$09} & Atoll & Yes & Yes & Yes\\ 
{\bf 4U~1728$-$34} & Atoll & Yes & Yes & Yes\\ 
{\bf 4U~1702$-$43} & Atoll & Yes & Yes & Yes\\ 
{\bf 4U~1820$-$30} & Atoll & Yes & Yes & Yes\\ 
{\bf Aql~X-1} & Atoll & Yes & Yes & Yes\\ 
{\bf 4U~1735$-$44} & Atoll & Yes & Yes & Yes\\ 
{\bf XTE~J1739$-$285} & Burster & Yes & Yes & Yes\\
{\bf EXO~1745$-$248} & Burster & Yes & Yes & Yes\\
{\bf 4U~1705$-$44} & Atoll & Yes & Yes & Yes\\ 
{\bf SAXJ1748.9$-$2021} & Atoll & Yes & Yes & Yes\\
{\bf IGR~J17191$-$2821} & Atoll & Yes & Yes & Yes\\
{\bf 4U~1915$-$05} & Atoll & Yes & Yes & Yes\\ 
SAXJ1808.4$-$3658 & AMP & Yes & Yes & Yes\\
SAXJ1750.8$-$2900 & Atoll  & Yes & Yes & Yes\\ 
RXJ1709.5$-$2639 & Other & Yes & Yes & Yes\\
XTE~J2123$-$058 & Atoll & Yes & Yes & No\\ 
KS~1731$-$260 & Burster & Yes & Yes & No\\
Cyg~X-2 & Z & Yes & Yes & No\\
EXO~0748$-$676 & Transient & Yes & Yes & No\\
MXB~1659$-$298 & Transient & Yes & Yes & No\\
XTE~J1723$-$376 & Burster & Yes & Yes & No\\
2S~0918-549 & Burster & Yes & Yes & No\\ 
1A~1246$-$588 & Burster & Yes & Yes & No\\
2A~1822$-$371 & AMP & Yes & Yes & No\\ 
1A~1744$-$361& Atoll & Yes & Yes & No\\
MXB~1730$-$33 & Burster & Yes & Yes & No\\
GRO~J1744$-$28 & AMP & Yes & Yes & No\\ 
Sco~X-1 & Z & Yes & No & No\\
GX~5$-$1 & Z & Yes & No & No\\
GX~17$+$2 & Z & Yes & No & No\\
GX~340$+$0 & Z & Yes & No & No\\
GX~349$+$2 & Z & Yes & No & No\\
XTE~J1751$-$305 & AMP & Yes & No & No\\
XTE~J1807$-$294 & AMP & Yes & No & No\\
XTE~J1814$-$338 & AMP & Yes & No & No\\
2S~1803$-$245 & Atoll & Yes & No & No\\
IGR~J18245$-$2452 & AMP & Yes & No & No\\
Ser~X-1 & Atoll & Yes & No & No\\
XTE~J1701$-$462 & Z & Yes & No & No\\
XTE~J0929$-$314 & AMP & No & No & No\\
4U~1746$-$37 & Atoll & No & No & No

\enddata
\tablenotetext{1}{kHz QPOs detected means detection using the analysis method in this paper and does not imply there are not detectable kHz QPOs in these sources.}
\label{tab:source_list}	
\end{deluxetable*}

Previous studies have shown the fractional rms amplitude of kHz QPOs increases steadily from $\sim$3 keV to 12 keV in nearly all sources \citep[see e.g.,][]{Berger_96,Wijnands_97,Zhang_96,Mendez_01} suggesting a source with thermal Comptonization rather than soft, blackbody emission.   Additionally, \citet{Gilf_03} and \citet{Rev_06} showed that the rms energy spectrum of neutron star LMXBs are consistent with a thermal Comptonization spectrum, which they suggest is due to the neutron star boundary layer (BL). Conservative arguments place the emission power of the BL as at least half of the total accretion powered emission (e.g., \citealt{Sunyaev_86}, and see \citealt{Gilf_14} for a review of neutron star BL physics).   Specifically, \citet{Gilf_03} showed that the rms energy spectrum of kHz QPOs in 4U~1608$-$52 is consistent with Comptonized emission.  Additionally, \citet{Peille_15} (for 4U~1608$-$52 and 4U~1728$-$34) and \citet{Troyer_17} (for Aql~X~-~1) both showed the energy-dependent covariance spectra (see Section~\ref{covariance}) are also consistent with Comptonized emission.  Thus, kHz QPOs are most likely associated with thermal Comptonization.

Several physical scenarios can lead to energy-dependent time lags.  For instance, where thermal accretion disk photons are Compton up-scattered by a hot corona or boundary layer to higher energies, delays between variations of higher-energy photons and variations of lower-energy photons are expected to arise.   This is because higher energy photons must have undergone more Compton scatterings, and hence spend a longer time within the corona compared to lower energy photons which undergo fewer scatterings.  Detailed analysis of Comptonization models suggest that this delay should scale with $\sim \log\frac{E_{high}}{E_{low}}$ \citep{Nowak_96}, which implies that higher energy photon oscillations occur after lower energy oscillations.  See \citet{Lee_Miller_98, Lee_01, Kumar_Misra_14} and \citet{Kumar_Misra_16} for detailed discussions of Comptonization lags in QPOs.  Study of kHz QPOs have shown that in some instances high energy oscillations lead the low-energy oscillations and in some instances that the reverse is true \citep[see e.g.,][]{Vaughan_98,Kaaret_99,deAvellar_13, Barret_13,Peille_15,Troyer_17}. 

Another process which could lead to time lags is reverberation \citep[see][for a review of X-ray reverberation in accreting black holes]{Uttley_14}.  Neutron star LMXBs often show prominent broad Fe~K$\alpha$ emission lines \citep[see, e.g.][]{cackett_10}. These lines are produced by irradiation of the accretion disk by high energy photons from the corona and/or boundary layer and are reflected and reprocessed by the accretion disk.  Since photons that arrive directly to the observer from the corona/boundary layer travel a shorter distance than those reflected off the accretion disk, a time lag should be present between the direct and reflected emission.

The first time lags in kHz QPOs were observed by \citet{Vaughan_98} in 4U~1608$-$52 and \citet{Kaaret_99} in 4U~1636$-$53 and used 3--4 energy bins.  More recently, spectral-timing analysis has been employed using significantly more energy bins and detailed studies of four neutron star LMXBs have been performed:  4U 1608-52 \citep{deAvellar_13, Barret_13}; 4U~1636$-$53 \citep{deAvellar_13, deAvellar_16}; 4U~1728$-$34 \citep{Peille_15}, and Aql~X~-~1 \citep{Troyer_17}.   These studies have shown for the lower kHz QPOs that the higher energy photons systematically arrive first.  The magnitude of these broadband lags have all been on the order of the light-travel time from the neutron star/boundary layer to the inner accretion disk.  

In cases where the upper kHz QPO is detected, the broadband lags are generally consistent with zero or exhibit a slight increase with energy \citep[see e.g.,][]{deAvellar_13,Peille_15}.  The interpretation of time lags is unclear and the complete picture of the source of time lags is likely complex, involving a combination of several physical processes as previously stated.  Understanding the ubiquity of time lags among different neutron star sources, as well as additionally complementary relationships between time lags and other spectral-timing products, may lead to further insights regarding their origin.    

To date, no comprehensive study exists that seeks to compute spectral-timing products of a large collection of neutron star LMXB systems. In this paper, we present the first systematic study of neutron star LMXB systems for data in the {\it RXTE} Proportional Counter Array (PCA) archive that looks at: (a) time lags as a function of QPO frequency and energy, (b) intrinsic coherence, and (c) fractional rms and covariance in relative rms units.   

In Section~\ref{data_reduction}, we describe the data reduction and kHz QPO search methodology.  In Section~\ref{QPO_class}, we discuss how we classify a detected QPO as either a lower or an upper.  In Section~\ref{spectral_timing} we detail the computational process and results of various spectral-timing products.  In Section~\ref{discussion} we discuss the implications and interpretation of our results, as well as some of the complementary nature of the various spectral-timing results.  Finally, in Section~\ref{conclusion} we summarize our results and state any conclusions that might be drawn from this work.

\section{Data Reduction} \label{data_reduction}

We searched the entire {\it RXTE/PCA} archive for event mode observations of neutron star LMXB systems.  In order to provide meaningful and consistent spectral-timing products, we considered only observations where the full 64 channel {\it RXTE/PCA} energy bands were available.  We also required observations with as good or better than 125~$\mu s$ time resolution.  The list of neutron star LMXB sources we considered as well as the nature of the observational limitations present in the {\it RXTE/PCA} archive are shown in Table~\ref{tab:source_list}.  In a small number of cases objects present in the archive did not have {\it event} mode data available.    

Once observations meeting the above criteria were identified, we created their associated response matrices using the FTOOL \textmyfont{pcarsp} to determine the channel-to-energy conversion table.  Barycentric correction was then performed using the FTOOL \textmyfont{faxbary}. We applied good time intervals (GTI) to avoid the Earth's limb, South Atlantic Anomaly (SAA), and to remove any thermo-nuclear bursts that occurred. 

We computed the discrete Fourier transform (DFT) over 4~s intervals, with a Nyquist frequency of 2048~Hz, and averaged 256 of these 4~s DFTs to produce a 1024~s average spectra.  This average time was adopted to facilitate easier detection of the upper kHz QPOs.  We binned the average power spectrum in 1~Hz bins.  The power spectrum was initially computed over the 3 -- 30~keV energy range.  We searched the power spectra for excess power between 300 -- 1300 Hz and fit a single Lorentzian profile using the maximum likelihood estimator (MLE) method as described in \citet{Barret_12}. We attempted a fit with a double Lorentzian profile in cases where a second excess power was found. In this way, each 1024~s power spectrum was characterized by one, two, or no candidate kHz QPOs.  Only individual 1024~s spectra where the QPO was significantly detected were considered for further analysis.  We consider a QPO as significant if the ratio of the fitted Lorentzian normalization ($I_{lor}$) to the error in the normalization($\Delta I_{lor}$) is: $I_{lor}/\Delta I_{lor} \ge$ 3.0 \citep[e.g.,][]{Bout_10}.  A significant QPO spectrum is a 1024 s power spectrum in which a kHz QPO is significantly detected.

The initial detections of kHz QPOs in the sources we analyze are well-documented in \citet{vanderKlis_00} with the following exceptions: \citet{XTEJ1739} for XTEJ1739$-$285, \citet{EXO1745} for EXO1745$-$248, \citet{SAXJ1748} for SAXJ1748.9$-$2021, and \citet{IGRJ17191} for IGR J17191$-$2821.

Within a single event mode file, each 1024 s spectra with a significantly detected QPO was shifted-and-added (see Méndez et al. 1998) to the mean QPO frequency. In cases where we detect both an upper and lower kHz QPO in a single 1024 s spectrum, the lowest frequency QPO was used for shifting and adding.  In all cases, the QPO frequency of each individual 1024 s spectrum within an event file was within 25 Hz of the mean QPO frequency.  This narrow frequency range demonstrates that we are not mixing lower and upper kHz QPOs with this shift-and-add procedure.  The resulting spectrum was again fit with a single or double Lorentzian profile using the MLE.  We used these profiles to classify the QPOs of each event file.  This provided cleaner profiles with higher S/N over individual spectra while still maintaining temporal continuity of the observation.  

kHz QPOs are not detected in all sources.  We emphasize that we only searched {\it event mode} observations with the previously mentioned constraints for kHz QPOs and attempted no detailed reconstruction of the QPO.  Additionally, we only performed single detections on 1024 s observation segments, while other studies might average over longer periods of time to detect kHz QPOs.  Finally, in some cases, kHz QPOs were detected in only a small number of averaged spectra and thus too few data points were available for meaningful spectral-timing analysis.

\section{\MakeLowercase{k}H\MakeLowercase{z} QPO Classification} \label{QPO_class}

\begin{figure*}
\centering
\includegraphics[trim=0.1cm 0.1cm 0.2cm 0.1cm, clip,width=17cm]{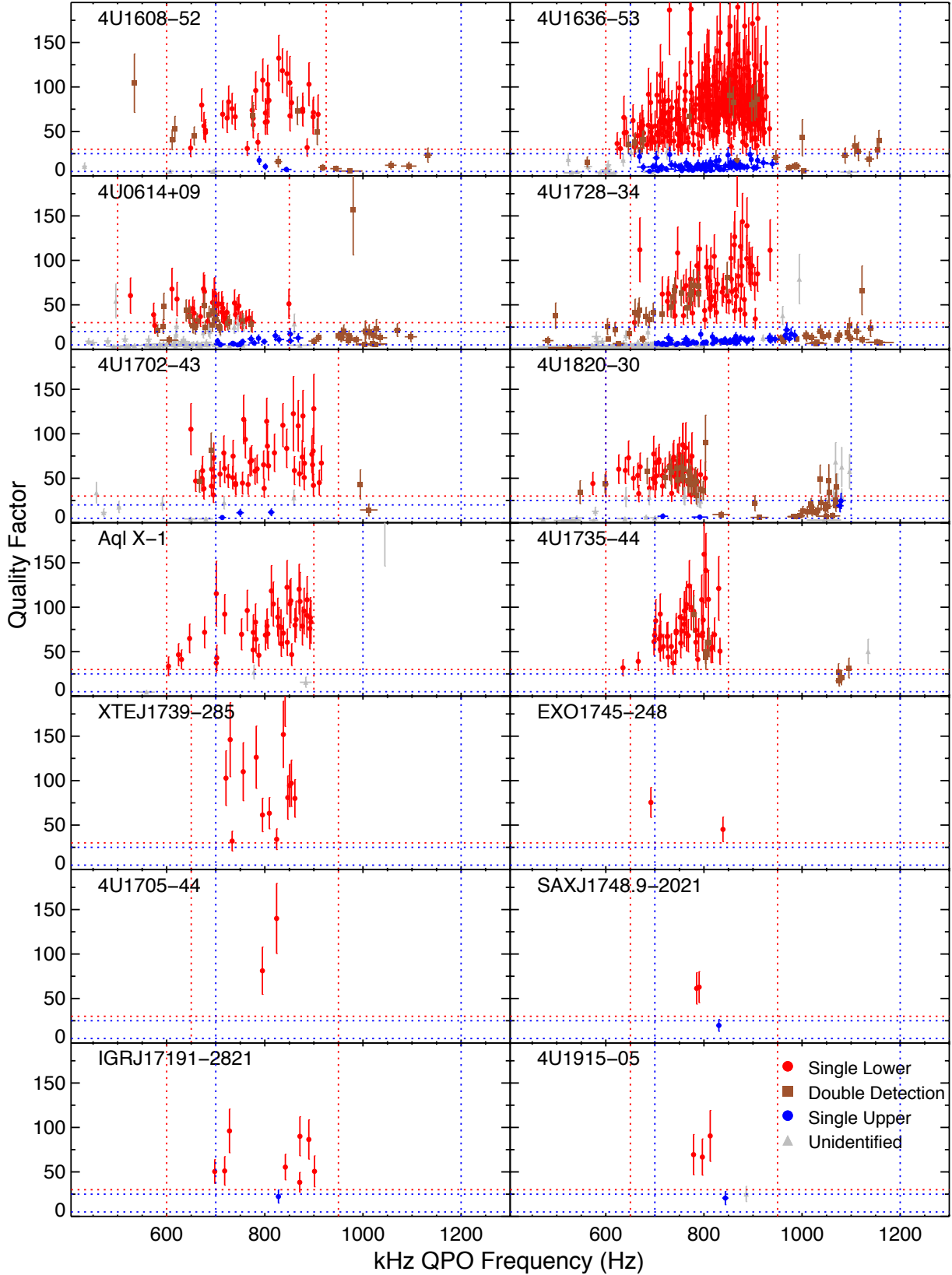}
\caption{Quality factor versus kHz QPO frequency for the (14) objects studied.  Quality factor is the ratio of the QPO frequency to the FWHM ($\nu/\Delta \nu$).   Each point represents a single shift-and-added event mode file using 1024 s spectra with a significantly detected kHz QPO.  The different quality factor tracks \citep[see e.g.,][]{Barret_06} are used to classify the QPOs as lower or upper in cases where the QPOs are ambiguous.  Limits in quality factor (Q) and frequency ($\nu$) were adopted from a visual-inspection of the data in the case of a single kHz QPO detection.  In all cases, a minimum quality factor of 5 was employed to ensure a good detection.  No maximum Q was used to filter detections of lower kHz QPOs.  The vertical red dashed lines indicate the frequency boundaries of the lower kHz QPOs.  The vertical blue dashed lines indicate the frequency boundaries of the upper kHz QPOs.  The horizontal red dashed line indicates the minimum quality factor for lower kHz QPOs.  The horizontal blue dashed line indicates the maximum quality factor for upper kHz QPOs.}
\label{fig:QF_plot}
\end{figure*}

We wish to partition the analysis products by QPO type (lower or upper kHz QPO).  Therefore, following \citet{Peille_15} in the case where a single kHz QPO is significantly detected in an event file, we classify that QPO using the quality factor (Q) parameter.  Quality factor is defined as the ratio of the QPO frequency to its full-width half-maximum (FWHM). In cases where two kHz QPOs are simultaneously detected, the quality factor is not needed as the classification is obvious. For a single kHz QPO detection, we look to the relationship between quality factor and QPO frequency.  The distinction between lower and upper kHz QPO can be generally inferred from this relationship. \citet{Barret_06} demonstrated that the lower and upper kHz QPOs follow distinct tracks on this diagram.  Using this relationship for each source, we set quality factor limits in order to classify single kHz QPO detections.  We compute the quality factor of the QPOs detected in each event file.  These plots are shown in Figure~\ref{fig:QF_plot}.  Looking at each object individually, we set frequency ranges for lower and upper kHz QPOs.  We also conservatively set allowed ranges of $Q$ to ensure unambiguous classification.

Once the kHz QPOs were classified, we were able to organize the event files for each object in order to combine observations with lower kHz QPOs and observations with upper kHz QPOs.  These event files were used in the computation of various spectral-timing products described in Section~\ref{spectral_timing}.  

\section{Spectral-Timing Products} \label{spectral_timing}

For each object, we combined all the event files for each QPO type across the entire {\it RXTE/PCA} archive.  This is necessary because in many cases the QPO is only weakly detected and spectral-timing analysis requires a significant number of counts in order to constrain the results.  To allow meaningful comparisons from source to source we perform an identical analysis on all objects. Having identified all kHz QPOs present in the data, we next produce a range of spectral-timing products of interest. See \citet{Nowak_99, Uttley_14} for a review of spectral-timing analysis methods.  We calculate all spectral-timing products following the prescription given in \citet{Uttley_14}.

The detector response varied over the multi-year range of data we averaged.  Therefore, it was necessary to rebin the channel-to-energy conversion tables to accommodate both small fluctuations in the energy boundaries of observations with identical channels and allow averaging of observations with different channel-to-energy conversion matrices.  All spectra were normalized to the total count rate using fractional rms normalization \citep{Miyamoto_91}.  We subtracted Fourier amplitudes where no source signal is nominally present (1350 Hz - 1700 Hz) from the cross-spectra in order to correct for dead time induced cross-talk \citep{vanderKlis_87}.

For the frequency-dependent lags, we compute the cross spectrum between the 3 -- 8 keV band and the 8 -- 20 keV for each 1024 s spectrum with a significantly detected kHz QPO, averaging over the FWHM of the QPO profile.  In this way, we obtain a lag measurement as a function of kHz QPO frequency.

For the energy-dependent spectral-timing products, we shifted-and-added \citep{Mendez_98} both the power spectra and the cross spectra of the reference band and the CI to the mean frequency of the classified QPO frequency range of each object.  We did this separately for the lower and upper kHz QPOs and only for spectra where a QPO was detected significantly as defined in Section~\ref{data_reduction}.  After shifting all the spectra, we fit a single Lorentzian to the shift-and-added spectra in order to characterize the FWHM of the shift-and-added QPO.   Averaging across one FWHM, we then computed the intrinsic coherence \citep{Vaughan_Nowak_97}, covariance \citep{Wilk_09} as well as the fractional rms, and energy-dependent lags \citep{Nowak_96,Vaughan_98}.  See \citet{Uttley_14} for details of each calculation.  The details of the intrinsic coherence results are shown in Section~\ref{coherence}, the rms and covariance results in Section~\ref{covariance}, a look at the frequency-dependence and energy-dependence of the lags are shown in Section~\ref{lag_freq} and \ref{lag_energy} respectively.  Unless otherwise stated, all errors reported are at the 1$\sigma$ level.

\subsection{Intrinsic Coherence} \label{coherence}

\begin{figure*}
\centering
\includegraphics[trim=0.1cm 0.1cm 0.2cm 0.1cm, clip,width=17cm]{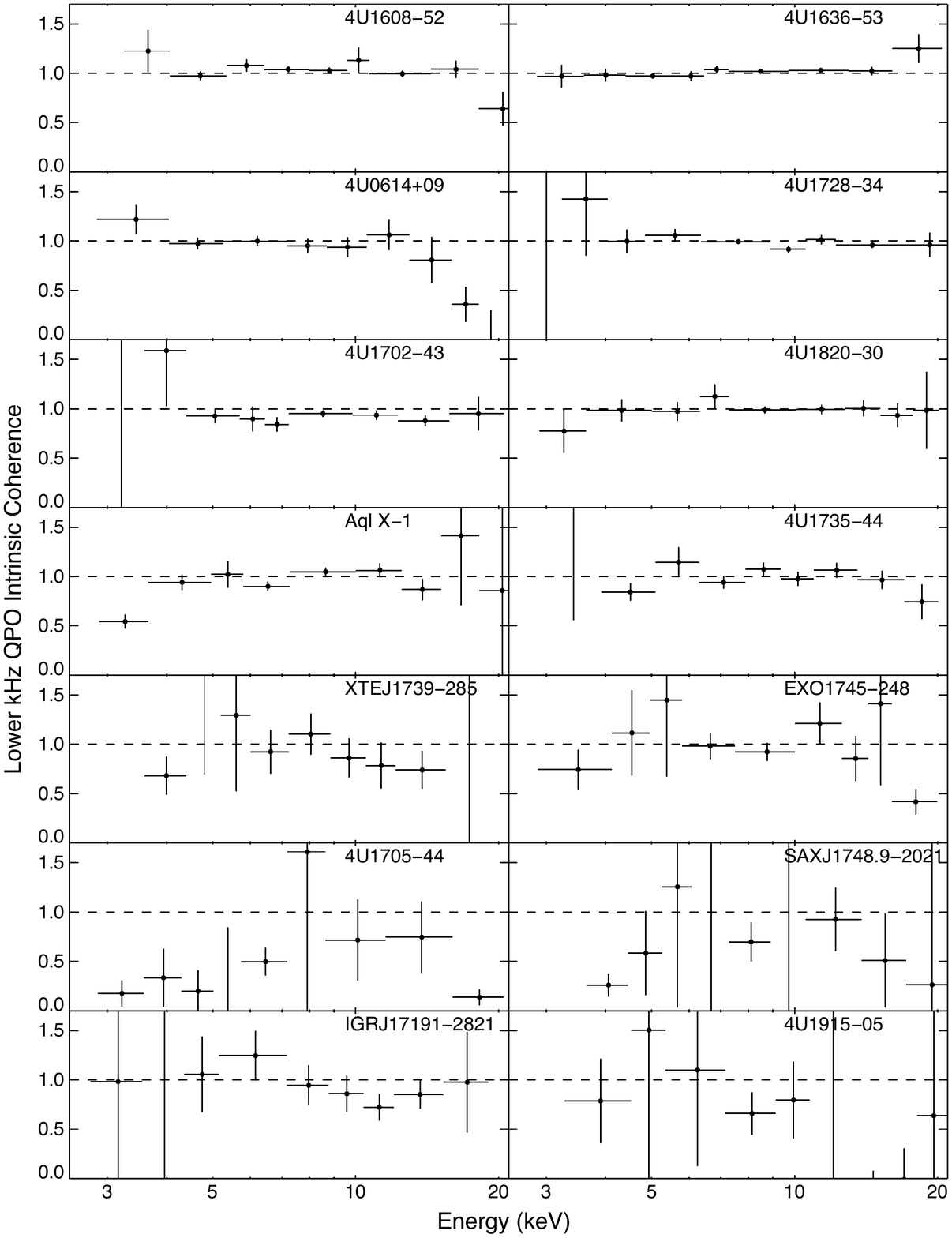}
\caption{Average intrinsic coherence as a function of energy for the lower kHz QPOs.  Intrinsic coherence measures the degree to which two signals are related by linear transformation.  In this case, one signal is the CI and the other is the reference band.  In general, objects with larger numbers of observations tend to have a more well-constrained coherence.  This is likely simply due to more signal available. }
\label{fig:lag_coh_lower}
\end{figure*}

\begin{figure*}
\centering
\includegraphics[trim=0.1cm 14.8cm 0.2cm 0.1cm, clip,width=17cm]{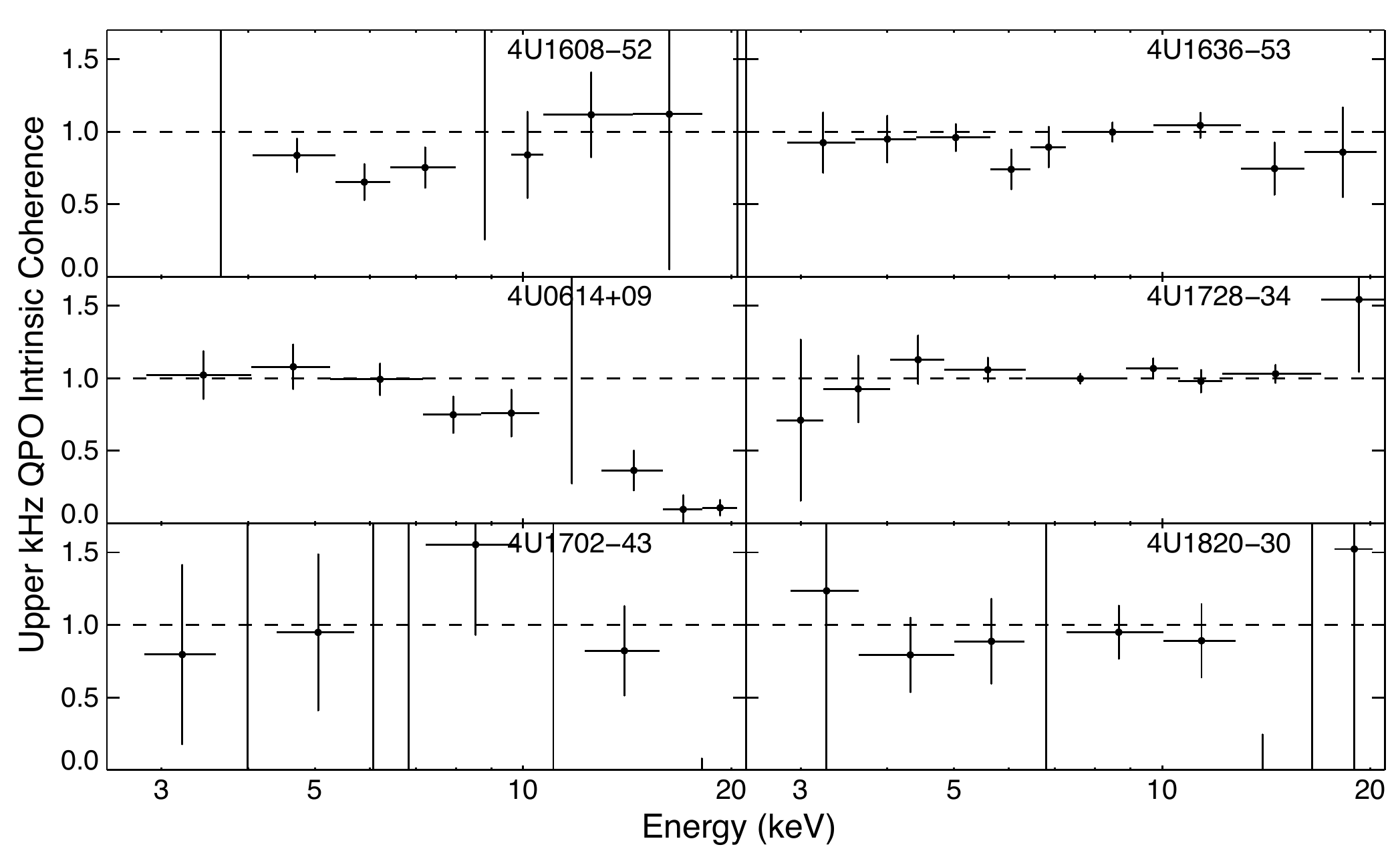}
\caption{Average intrinsic coherence as a function of energy for the upper kHz QPOs.  See the caption in Figure \ref{fig:lag_coh_lower} for additional details.}
\label{fig:lag_coh_upper}
\end{figure*}

The intrinsic coherence provides a quantitative measure of how one signal is related to another by linear transformation \citep[e.g.,][]{Vaughan_Nowak_97}.  Generally speaking, if we wish to interpret the various spectral-timing products, especially time lags, as a result of some type of physical process, i.e. energy-dependent Comptonization delays or light-travel time delays due to spatial separation of one or more emitting regions, it is crucial that the underlying signals have near unity coherence.  In other words if intrinsic coherence is not near unity we lose the underlying attachment of the two signals to a direct physical process.  Thus, understanding the coherence of the kHz QPOs across the energy range of interest is an important place to start.  Both \citet{Vaughan_98} and \citet{Kaaret_99} showed that kHz QPOs in 4U~1608$-$52 and 4U~1636$-$53 have high (near unity) coherence. Additionally, \citet{deAvellar_13} nicely showed the frequency dependence of the coherence across the kHz QPO profile.  In Figure~1 of that paper, the unity coherence of the QPOs is illustrated.

The average coherence as a function of energy for the lower and upper kHz QPOs are shown in Figures~\ref{fig:lag_coh_lower} and \ref{fig:lag_coh_upper} respectively.  Here, we adopt confidence limits for the high-signal, high-coherence regime as defined in \citet{Nowak_96} for all values of coherence due to the difficulties in quantifying the coherence confidence limits outside of this regime \citep[e.g.,][]{Vaughan_Nowak_97}.  Thus, confidence limits for low values of coherence should be considered unreliable.

General speaking, we find that both kHz QPOs show near-unity coherence across a majority of the energy range we use.  Objects with larger amounts of data generally show more well-constrained coherence.  Additionally, at higher energies, the QPO signal can become dominated by noise, seen most convincingly in 4U~0614$+$09, which can reduce the coherence.

\subsection{Covariance} \label{covariance}

The covariance measures the variability power in an energy bin that is correlated with the reference band \citep[see][]{Wilk_09,Uttley_11,Uttley_14}.  For high coherence signals, the covariance is essentially the rms spectrum with intrinsically higher S/N \citep{Uttley_14}.  It is well known that the rms spectrum of kHz QPOs increases with increasing energy up to about $\simeq$15 keV, where it levels off \citep[see e.g.,][]{Berger_96,Zhang_96,Wijnands_97,Mendez_01}.  \citet{Peille_15} showed similar behavior in the fractional covariance for 4U~1728$-$34, and this trend was recently extended below 3 keV through {\it NICER} observations of an upper kHz QPO in 4U~0614+09 \citep{bult18}.  Our results are generally consistent with this behavior of increasing fractional covariance with energy, flattening off above 15 keV.  

\begin{figure*}
\centering
\includegraphics[trim=0.1cm 0.1cm 0.2cm 0.1cm, clip,width=17cm]{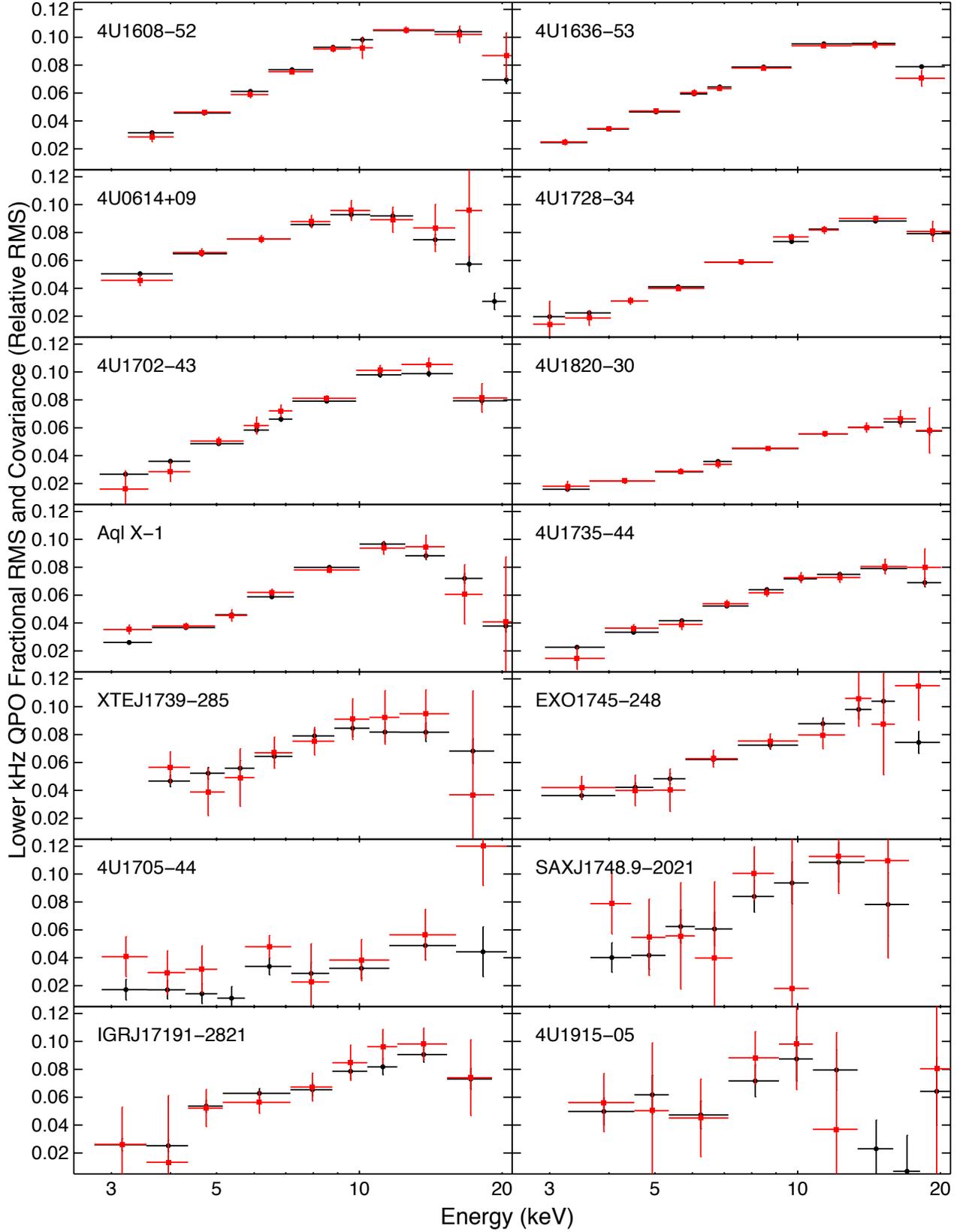}
\caption{Average fractional rms and covariance in fraction rms units as a function of energy for the lower kHz QPOs.  Fractional rms is shown in red squares and covariance in black dots.  In general, objects show increasing fractional variability with increasing energy.  The divergence of fractional rms and covariance implies a drop in coherence, seen here at the lowest energy bins and the highest.  This behavior is consistent with the behavior of the coherence seen in Figure~\ref{fig:lag_coh_lower}.  At energies above $\simeq 15$ keV,  the covariance falls off generally, while the fraction rms levels off or increases while becoming less well-constrained. The drop in covariance at high energies generally tracks the drop in intrinsic coherence which reduces the degree to which one can interpret the covariance as conveying the variability of the QPO.  Poorly constrained rms values due to poor S/N are not shown.}
\label{fig:lag_cov_lower}
\end{figure*}

\begin{figure*}
\centering
\includegraphics[trim=0.1cm 14.8cm 0.2cm 0.1cm, clip,width=17cm]{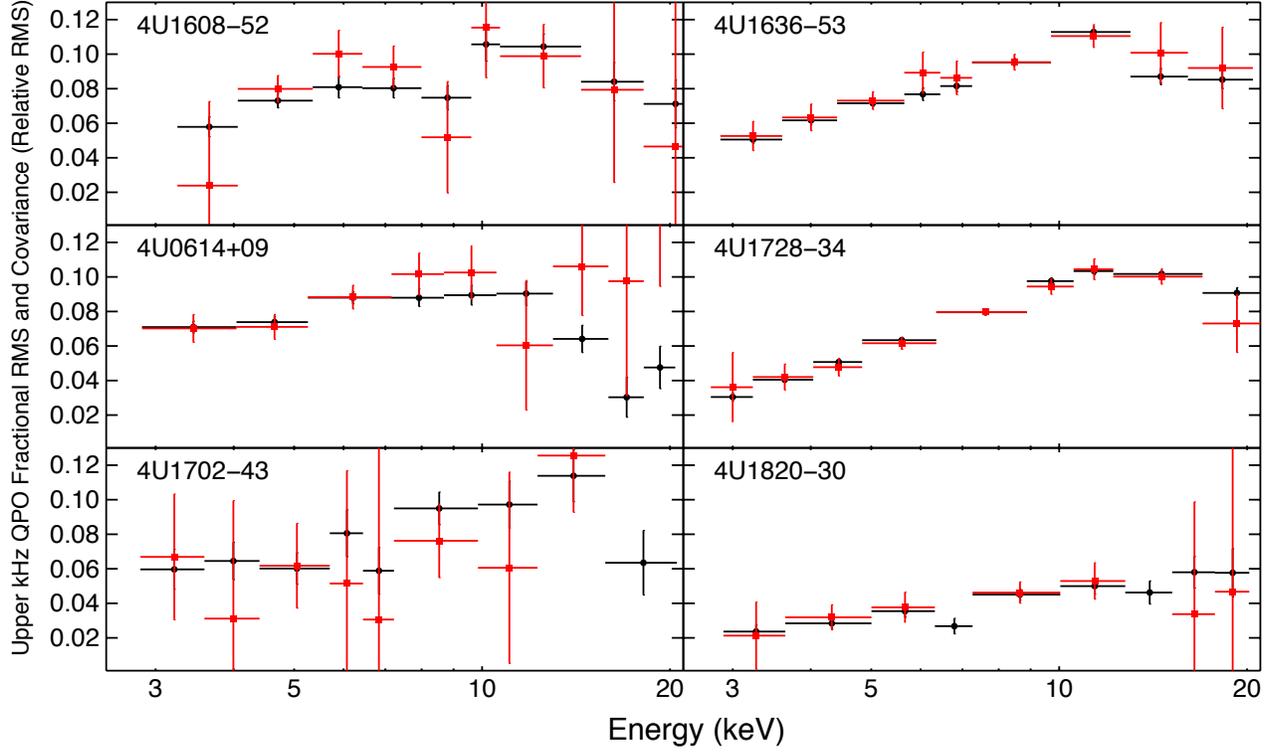}
\caption{Average fractional rms and covariance as a function of energy for the upper kHz QPOs.  The fractional rms is shown in red squares and covariance in black dots. Objects generally show increasing variability with increasing energy, with the behavior of upper kHz QPOs similar to that of the lower kHz QPOs.}
\label{fig:lag_cov_upper}
\end{figure*}

In order to illustrate the usefulness and limitations of the covariance statistic, we have computed both the rms and covariance in fractional rms units.  These data for the lower and upper kHz QPOs are shown in Figures~\ref{fig:lag_cov_lower} and \ref{fig:lag_cov_upper} respectively.  The errors on the rms and covariance calculation are valid in the limit where the intrinsic coherence is unity \citep{Uttley_14}.

In general, we note increasing rms and covariance with increasing energy for both the lower and upper kHz QPOs.  In a few cases, a drop in rms and covariance is seen in the highest energy bin.  In 4U 0614$+$09 we note decreasing rms and covariance at energies above $\sim$15 keV.  We also note that there is sometimes a divergence between the rms and covariance.  These are both likely due to a drop in coherence seen at those energies.  For example, see Figure~\ref{fig:lag_coh_lower}.  At higher energies, the QPO signal can become dominated by noise, which can result in the reduction of both the rms and covariance.

\subsection{Lags vs. Frequency} \label{lag_freq}

\begin{figure*}
\centering
\includegraphics[trim=0.1cm 0.1cm 0.2cm 0.1cm, clip,width=17cm]{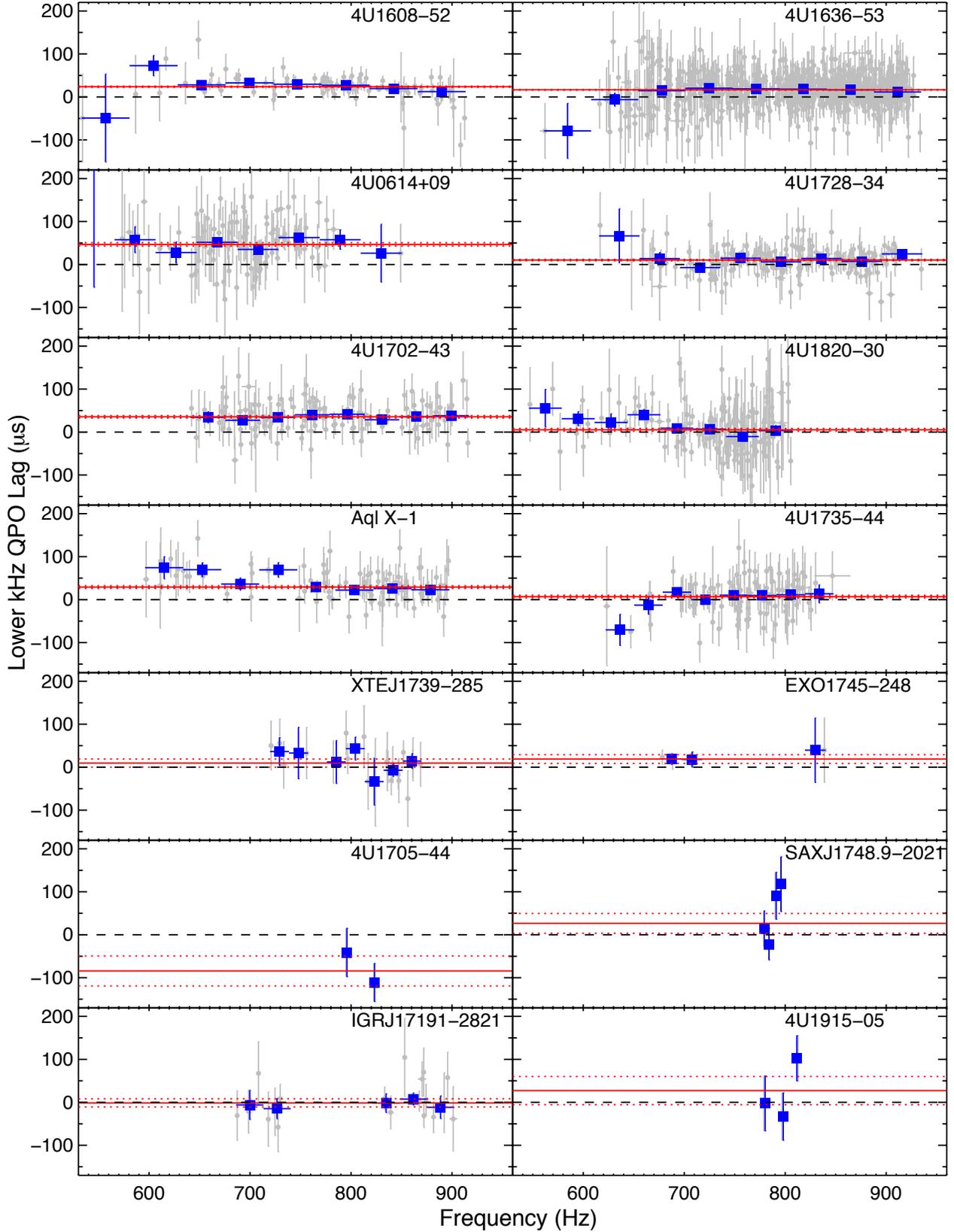}
\caption{Broad energy lags as a function of kHz QPO frequency for the lower kHz QPO.  The low energy band is 3 -- 8 keV and the high energy band is 8 -- 20 keV.  Each small dot represents a single 1024 s spectrum with a significantly detected lower kHz QPO.  The data are binned in a maximum of eight equally-spaced frequency bins shown in large blue squares.  The average lag with 1$\sigma$ error is shown in red and the values of the average lags are shown in Table~\ref{tab:lagfreq_lower}.  A positive lag indicates that variability in the higher energy band (8 -- 20 keV) arrives before variability in the lower energy band (3 -- 8 keV).  In general, lower kHz QPOs show positive average lags.  In all cases, there is either a slight decrease in lags as a function of QPO frequency or no significant frequency dependence.  In objects with more than (5) data points, the linear fit parameters are also shown in Table~\ref{tab:lagfreq_lower}.}
\label{fig:lag_freq_lower}
\end{figure*}

\begin{figure*}
\centering
\includegraphics[trim=0.1cm 14.8cm 0.2cm 0.1cm, clip,width=17cm]{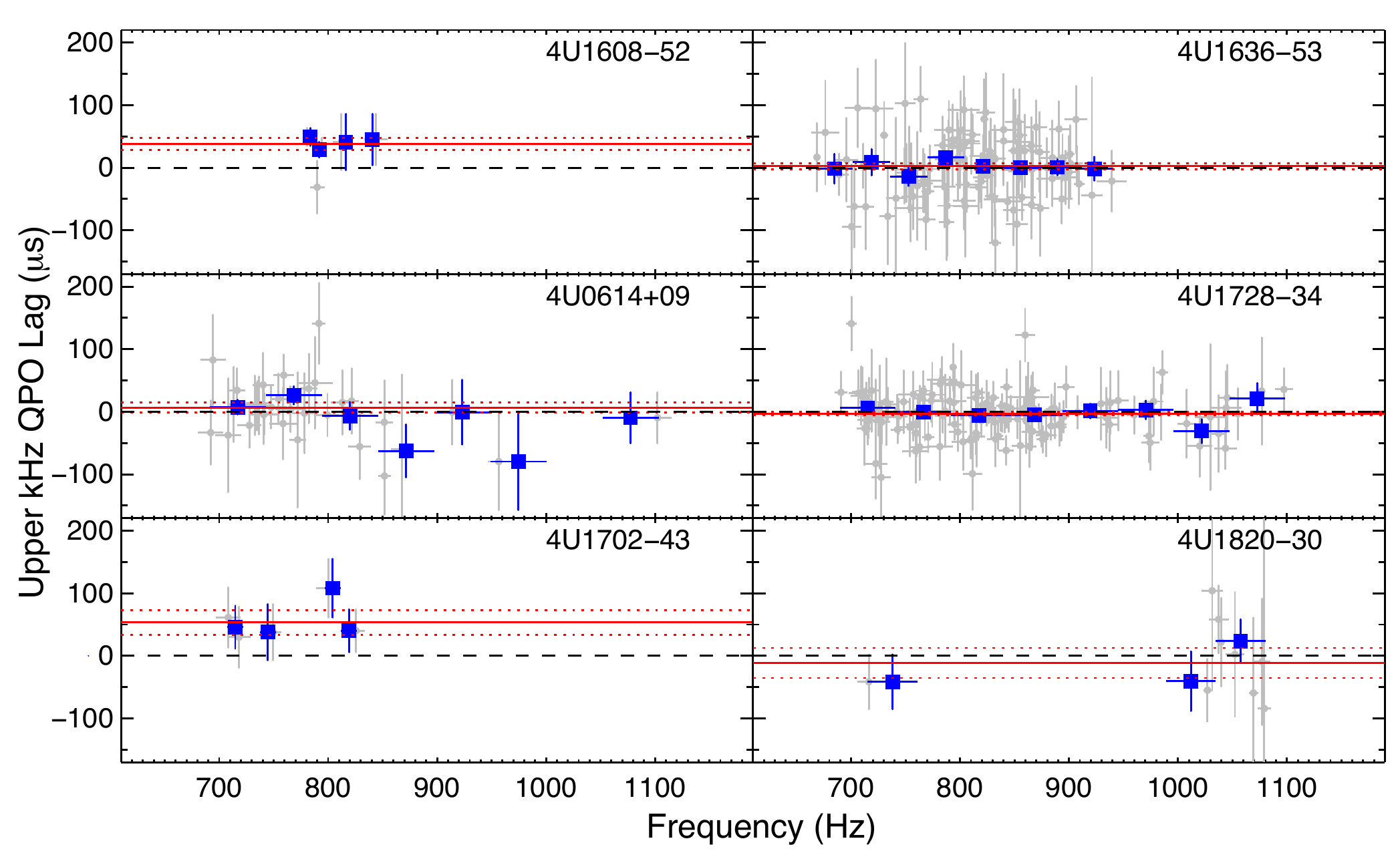}
\caption{Broad energy lags as a function of kHz QPO frequency for the upper kHz QPO.  See the caption of Figure \ref{fig:lag_freq_lower} for additional details.  The average lag with 1$\sigma$ error is shown in red and the values of the average lags are shown in Table~\ref{tab:lagfreq_upper}.  The upper kHz QPOs show lags which are consistent with 0 except in the cases where the number of 1024 s segments with lags is small, i.e. 4U~1608$-$52 and 4U~1702$-$43.  In all cases, there is no significant dependence of the lag on frequency.  The linear fit parameters are also shown in Table~\ref{tab:lagfreq_upper}.}
\label{fig:lag_freq_upper}
\end{figure*}
\begin{deluxetable*}{ccccccc}
\tablewidth{0pt}
\tablecolumns{7}
\tabletypesize{\small}
\tablecaption{Lower kHz QPO Lag/Frequency Data.  This includes: the weighted mean of the lags, the best fit slope and intercept of the linear model we fit to the unbinned data, the $\chi^{2}$(dof) of both the linear model and the constant lag model, and the result of an F-test indicating which of these two models is preferred and the statistical significance of the conclusion.  In the cases of EXO~1745$-$248, 4U~1705$-$44, SAXJ1748.9-2021, and 4U~1915$-$05, we assess the number of data points ($< 5$) too small to provide meaningful $\chi^{2}$(dof) analysis.}
\tablehead{
\colhead{Object ID} & \colhead{Average Lag ($\mu s$)} & \colhead{Slope ($\mu s/Hz$)} & \colhead{Intercept ($\mu s$)}  & \colhead{$\chi^{2}$(dof) constant} & \colhead{$\chi^{2}$(dof) linear} & \colhead{Favored Model}}
\startdata
4U~1608$-$52 & 23.7 $\pm$ 1.3 & $-0.09 \pm$ 0.02 & 96.3 $\pm$ 1.3  & 115.0(92) & 97.9(91) & Linear 3.8$\sigma$ \\ 
4U~1636$-$53 & 16.5 $\pm$ 0.8 & $-0.02~\pm$ 0.02 & 36.0 $\pm$ 0.8  & 767.3(742) & 765.3(741)& Linear 1.4$\sigma$\\ 
4U~0614$+$09 & 46.8 $\pm$ 4.0 & 0.07 $\pm$ 0.1 & $-0.8~\pm$ 4.1  & 117.4(112) & 116.9(111) & Linear 0.69$\sigma$\\ 
4U~1728$-$34 & 9.5 $\pm$ 1.6 & 0.01 $\pm$ 0.03 & 1.9 $\pm$ 1.6  & 157.4(175) & 156.8(174) & Linear 0.81$\sigma$\\
4U~1702$-$43 & 35.6 $\pm$ 2.8 & 0.03 $\pm$ 0.04 & 14.7 $\pm$ 2.9  & 112.3(114) & 111.9(113) & Constant 0.63$\sigma$\\
4U~1820$-$30 & 5.4 $\pm$ 3.1 & $-0.26~\pm$ 0.07 & 193 $\pm$ 3  & 175.2(153) & 162.0(152) & Linear 3.4$\sigma$\\ 
Aql~X-1 & 29.2 $\pm$ 3.0 & $-0.15~\pm$ 0.04 & 154 $\pm$ 3 & 96.2(88) & 86.1(87)  & Linear 3.1$\sigma$\\
4U~1735$-$44 & 6.9 $\pm$ 3.1 & 0.12 $\pm$ 0.08 & $-82.5~\pm$ 3.2  & 108.9(109) & 106.7(108) & Linear 1.5$\sigma$\\ 
XTEJ1739$-$285 & 9.6 $\pm$ 9.3 & $-0.3~\pm$ 0.2 & 267 $\pm$ 10  & 10.0(21) & 8.3(20) & Linear 1.9$\sigma$\\
EXO~1745$-$248 & 18.9 $\pm$ 10.1 & 0.06$\pm$ 0.43 & $-22.0~\pm$ 11.9 & 0.2(3) & 0.2(2) & - \\ 
4U~1705$-$44 & $-84.3~\pm$ 35.0 & $-2.4\pm$ 2.5 & 1875~$\pm$ 44 & 0.9(1) &  0(0) & - \\ 
SAXJ1748.9$-$2021 & 26.6 $\pm$ 23.0 & 8.7$\pm$ 4.8 & $-6842~\pm$ 31  & 5.2(3) &  1.7(2) & - \\  
IGRJ17191$-$2821 & $-1.4~\pm$ 9.5 & 3.4 $\pm$ 2.5 & $-73~\pm$ 10 & 11.4(18) & 11.0(17)  & Linear 0.77$\sigma$\\ 
4U~1915$-$05 & 27.2 $\pm$ 32.8 & 3.4$\pm$ 2.5 & $-2670~\pm$ 33  & 3.4(2) &  1.5(1) & -
\enddata
\label{tab:lagfreq_lower}
\end{deluxetable*}
\begin{deluxetable*}{ccccccc}
\tablewidth{0pt}
\tablecolumns{7}
\tabletypesize{\small}
\tablecaption{Upper kHz QPO Lag/Frequency Data.  For additional details, see the caption of Table \ref{tab:lagfreq_lower}.  For the upper kHz QPO, in the cases of 4U~1608$-$52 and 4U~1702$-$43 we assess the number of data points ($< 5$) too small to provide meaningful $\chi^{2}$(dof) analysis.}
\tablehead{
\colhead{Object ID} & \colhead{Average Lag ($\mu s$)} & \colhead{Slope ($\mu s/Hz$)} & \colhead{Intercept ($\mu s$)} &  \colhead{$\chi^{2}$(dof) constant}   & \colhead{$\chi^{2}$(dof) linear} & \colhead{Favored Model}}
\startdata
4U~1608$-$52 & 38.2 $\pm$ 9.3 & $-0.13~\pm$ 0.65 & 144 $\pm$ 9  & 3.3(4) & 3.2(3) & - \\ 
4U~1636$-$53 & 2.6 $\pm$ 4.8 & $-0.01~\pm$ 0.08 & 12.4 $\pm$ 4.8  & 81.9(103) & 81.9(102) & Equal\\ 
4U~0614$+$09 & 6.8 $\pm$ 8.3 & $-0.10~\pm$ 0.10 & 90.0 $\pm$ 8.5  & 18.1(32) & 16.9(31) &  Linear 1.4$\sigma$\\ 
4U~1728$-$34 & $-2.9~\pm$ 2.8 & $-0.04~\pm$ 0.04 & 30.0 $\pm$ 2.7  & 113.3(127) & 112.1(126) &  Linear 1.1$\sigma$\\
4U~1702$-$43 & 53.2 $\pm$ 19.6 & 0.10 $\pm$ 0.41 & $-25~\pm$ 21 & 1.9(4) & 1.8(3) & - \\ 
4U~1820$-$30 & $-11.4~\pm$ 24.0 & 0.13 $\pm$ 0.16 & $-136~\pm$ & 3.7(8) & 3.1(7)  &  Linear 1.1$\sigma$\ 
\enddata
\label{tab:lagfreq_upper}	
\end{deluxetable*}

To establish the presence of any lags, and explore any possible frequency dependence, we computed lags between two broad energy bins: 3 -- 8 keV and 8 -- 20 keV.  We note here that the broad--energy lags are not calculated using the channel-of-interest and reference band convention.  These lags are produced directly between the two broad energy bins.  We considered lower kHz QPOs separately from uppers.  We were able to derive lower kHz QPO lag vs frequency relationships for 14 objects and upper kHz QPO lag vs frequency relationships for 6 objects.  The lag vs frequency relationships for the lower kHz QPOs and the upper kHz QPOs are shown in Figures~\ref{fig:lag_freq_lower} and \ref{fig:lag_freq_upper} respectively.  A positive lag indicates that the higher-energy band photon variability arrives before the lower-energy band photon variability.  We binned the lag/frequency data into a maximum of eight bins in an effort to highlight any frequency dependence.

In general, the lags for the lower kHz QPOs show positive average lags, i.e. so-called `soft lags' where the lower energy photons lag the higher energy ones.  This result is seen in all previously published work where such lags have been computed \citep{Barret_13,deAvellar_13,Peille_15,deAvellar_16,Troyer_17}.  The upper kHz QPOs generally have reduced data quality.  In general, we find lags consistent with zero or slightly negative average lags, except in two cases.  Zero or slightly negative average lags are also consistent with previously published work and illustrates that the nature of the lags is fundamentally different between the lower and upper kHz QPO \citep{deAvellar_13,Peille_15,deAvellar_16,Troyer_17}.  Moreover, the values of the average lags are consistent with all comparable previous analyses which include: 4U~1608$-$52, 4U~1636$-$53, 4U~1728$-$34, and Aql~X-1.   Both 4U~1608$-$52 and 4U~1702$-$43 show positive average lags for the upper kHz QPOs.  In both cases, there are relatively few data and the upper kHz QPOs are very weakly detected as shown by their poorly constrained coherence.

While in some cases significant upper kHz QPOs are found in event file shift-and-added spectra shown in Figure~\ref{fig:QF_plot}, an individual 1024 s spectrum with a significant upper kHz QPO may not exist and thus not appear in the lag vs. frequency analysis. 

In general, the dependence of lag on kHz QPO frequency is likely complex.  \citet{Barret_13} showed the binned frequency-dependent lags for 4U 1608$-$52 increase monotonically to $\sim$~700~Hz and then decrease monotonically at higher frequencies.  \citet{deAvellar_13} showed no dependence of lags on frequency for 4U~1608$-$52 using a smaller data set than \citet{Barret_13}.  Additionally, \citet{deAvellar_13} showed a slight decrease in lower kHz QPO lags for 4U~1636$-$53 above 850~Hz with no frequency dependence for the upper kHz QPO. \citet{deAvellar_16}, who computed phase lags, find no significant frequency dependence for either the lower or upper kHz QPOs in 4U~1636$-$53.  When fitting the lag/frequency relationship, both \citet{deAvellar_13} and \citet{deAvellar_16} note that linear or quadratic fits are not favored over a constant fit.   In 4U~1728$-$34, \citet{Peille_15} find no frequency dependence for the lower kHz QPO lags and the possibility of a slight decrease of the lags at higher frequency for the upper kHz QPO. In Aql~X-1, \citet{Troyer_17} find no significant frequency dependence on the lags for the lower kHz QPO.

In all previous works, various energy boundaries, power spectral averaging times, and/or data sets are used.  \citet{Barret_13} uses a 128 s averaged power spectra to calculate the lags, and a smaller data set than this work.   Both \citet{deAvellar_13} and \citet{deAvellar_16} use different energy bin boundaries and also a smaller data set than this work as well as a 16 s power spectra average time.  While \citet{Peille_15} and \citet{Troyer_17} use the same energy boundaries and similar data sets, \citet{Troyer_17} uses a 256 s power spectra average time.

Our use of a 1024 s spectra averaging time, while necessary to maximize the kHz QPO detections in many sources, could mask the underlying lag/frequency relationship.  Thus we choose to apply a simple linear fit of the unbinned data in an effort to characterize the lag-frequency relationship, while noting the underlying relationship could be more complex.  While not perfect, this does allow for meaningful comparisons of any trends in the lag/frequency relationship between sources.   

In order to determine if a linear trend is favored over a constant, we computed the $\chi^{2}$ statistic for both a linear and constant model of the data and then used an F-test to determine the significance of the statistically favored model.  The average lag, best-fit slope and intercept, and $\chi^{2}$ values with the statistically favored model are shown in Table~\ref{tab:lagfreq_lower} for the lower kHz QPOs and Table~\ref{tab:lagfreq_upper} for the upper kHz QPOs.  For the lower kHz QPOs in four objects: EXO~1745$-$248, 4U~1705$-$44, SAXJ1748.9-2021, and 4U~1915$-$05, and the upper kHz QPOs in two objects: 4U~1608$-$52 and 4U~1702$-$43; we assessed the number of data points ($<5$) too small to provide a meaningful $\chi^{2}$ comparison of the data.

For the lower kHz QPOs, in three objects, 4U~1608$-$52, 4U~1820$-$30, and Aql~X-1, the linear model is favored with $>3\sigma$ significance over the constant model.  In all other models there is no statistically significant difference between a linear fit and a constant fit.  While the data do generally show lags that decrease slightly with frequency (though at low statistical significance), because of the paucity of high-quality data for most of the sources and the fact that splitting into multiple frequency bins spreads the number of photons available for analysis across multiple spectral-timing products, we chose to average over all frequencies for all energy-dependent  analyses (coherence, covariance and lag-energy spectra).  This allows for increased S/N when averaging all data together instead of splitting into multiple bins, and enables a systematic method of analysis across all sources.

\subsection{Energy-Dependent Lags} \label{lag_energy}

\begin{figure*}
\centering
\includegraphics[trim=0.1cm 0.1cm 0.2cm 0.1cm, clip,width=17cm]{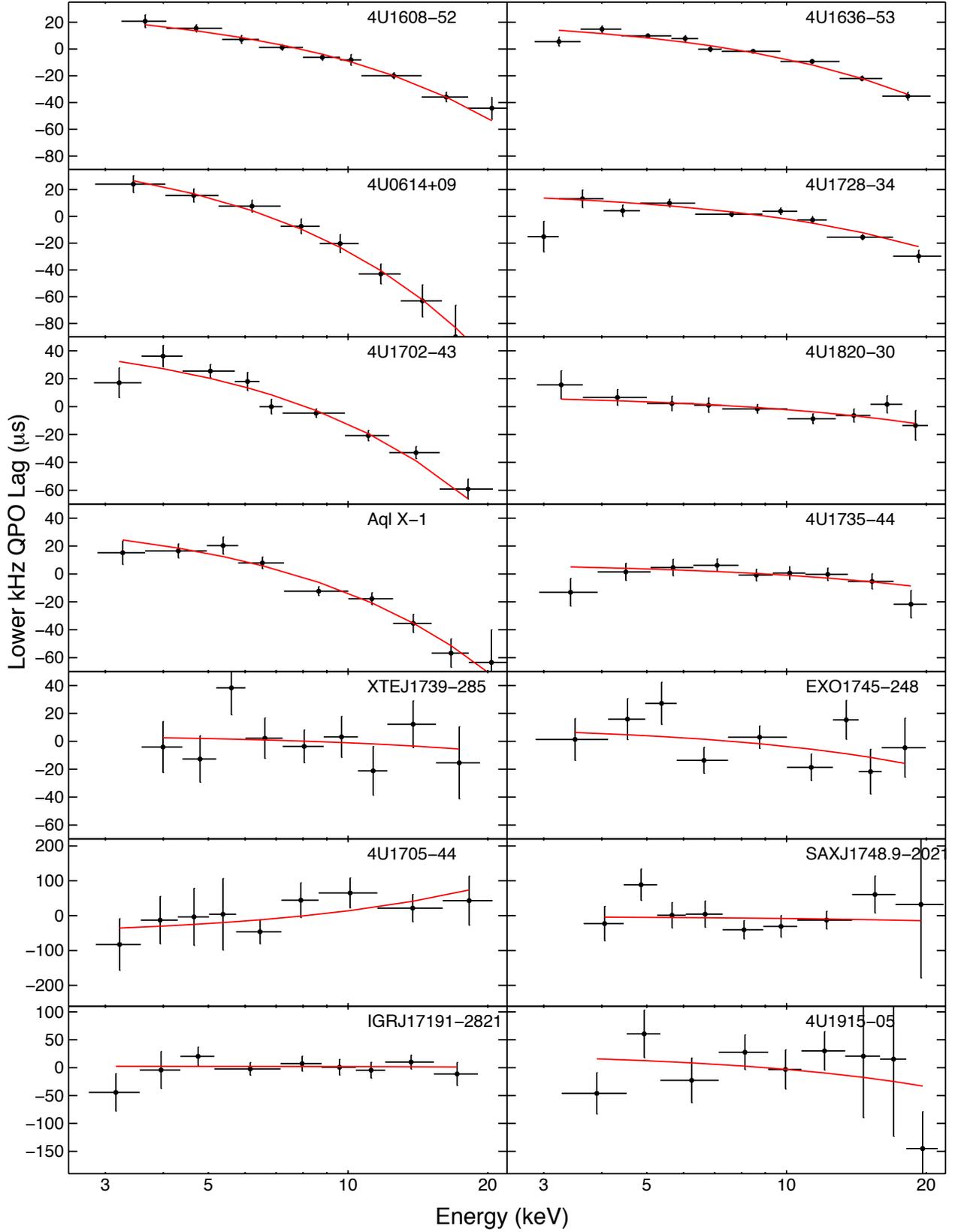}
\caption{Energy-dependent lags for the lower kHz QPO.  A positive value of lag in an energy bin indicates that the variability in that bin arrives after the variability in the reference band.   In general, lower kHz QPOs show lags with a decreasing trend as energy increases.  In order to further characterize the lags, we performed a linear fit in all cases.  The best-fit line is shown in red.  The best-fit intercept and slope with confidence limits are shown in Table~\ref{tab:lagen_lower}.}
\label{fig:lagen_lower}
\end{figure*}

\begin{figure*}
\centering
\includegraphics[trim=0.1cm 14.8cm 0.2cm 0.1cm, clip,width=17cm]{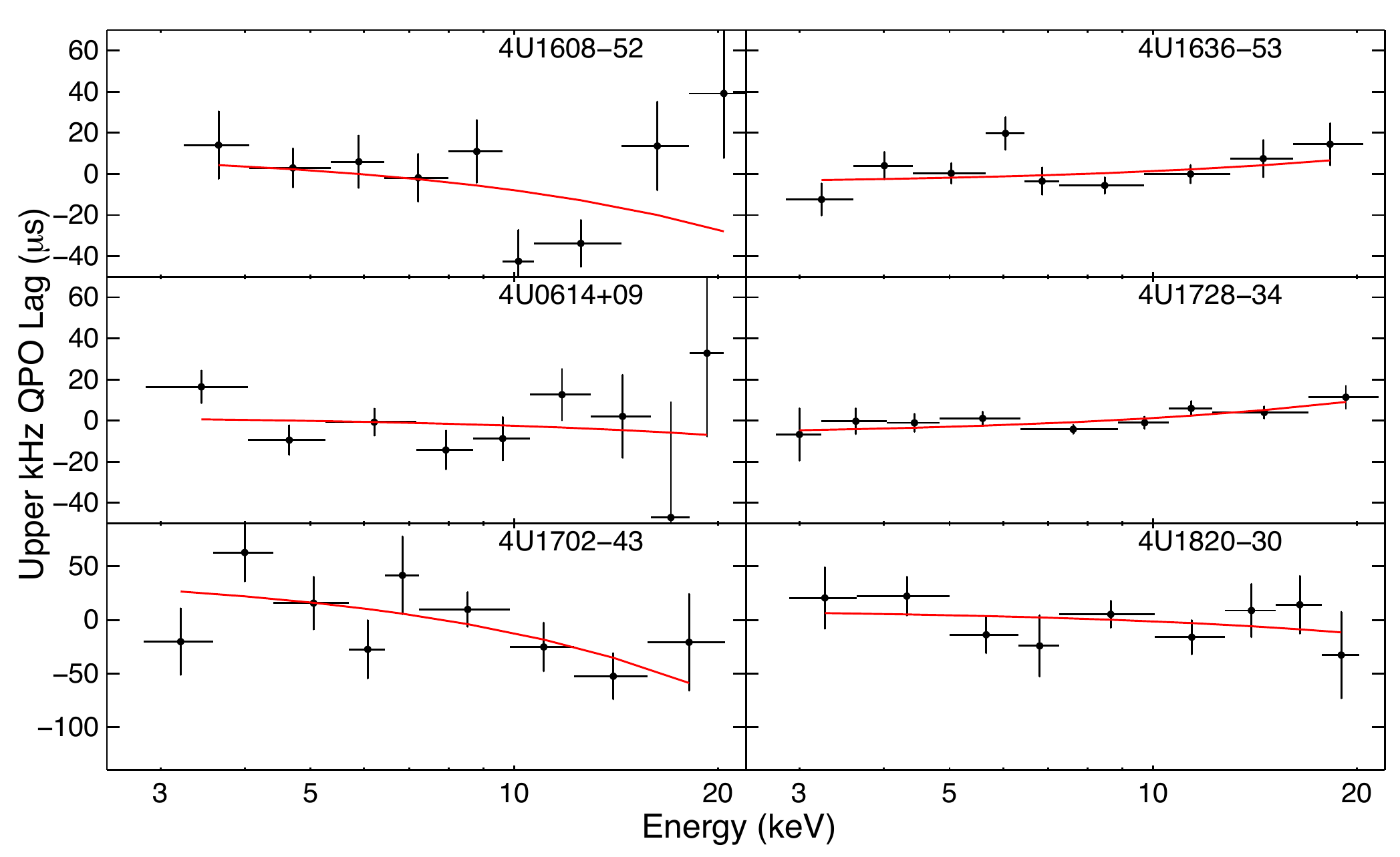}
\caption{Energy-dependent lags for the upper kHz QPO.  In general, upper kHz QPOs show lags which are consistent with zero lag.  In order to characterize the lags, we performed a linear fit in all cases.  The best-fit line is shown in red.  The best-fit intercept and slope with confidence limits are shown in Table~\ref{tab:lagen_upper}.} 
\label{fig:lagen_upper}
\end{figure*}

Energy-dependent lags offer a way to begin describing the role of various time-dependent physical process involved in the creation of kHz QPOs.  These data illustrate the time lag in each energy bin relative to the reference band.  They should be read as relative lags, with the most negative lag being the energy from which the photons arrive first.  We show lags as a function of energy for the lower and upper kHz QPOs in Figures~\ref{fig:lagen_lower} and \ref{fig:lagen_upper} respectively.  To further characterize the data, we performed linear fits and show the best-fit slope and intercept as well as then number of significant QPO spectra we averaged to obtain the lag/energy spectrum in Tables~\ref{tab:lagen_lower} and \ref{tab:lagen_upper} for the lower and upper kHz QPOs respectively.  The best-fit slope and intercept are used as a simple way to characterize the data.  Additionally, the shape of the lag/energy spectrum depends on the strengths of the components of the energy spectrum and thus would not necessarily fit a linear function.

For the lower kHz QPOs, the trend is generally monotonically decreasing lags with increasing energy.  The lag of each energy bin -- channel of interest (CI) --  is measured with respect to the CI-subtracted reference band energy range (3 -- 20 keV).  This trend shows that the variability at lower energies lags the variability at higher energies, with the highest energy photons arriving first. The upper kHz QPOs generally show lags that are flat and in some cases show an increase in the lags at higher energies albeit with poorer statistics than lower kHz QPO.  This is consistent with the results of \citet{deAvellar_13} and \citet{Peille_15}, which show different behavior of energy-dependent lags of the lower and upper kHz QPO for 4U~1608$-52$, 4U~1636$-53$, and 4U~1728$-34$.   This implies that the physical mechanisms responsible for the creation of lower and upper kHz QPOs are likely different.

\begin{deluxetable*}{cccc}
\tablewidth{0pt}
\tablecolumns{4}
\tabletypesize{\small}
\tablecaption{Lower kHz QPO Lag/Energy Data.}
\tablehead{
\colhead{Object ID} & \colhead{Number of Significant QPOs} & \colhead{Best Fit Slope ($\mu s/keV$)} & \colhead{Best Fit Intercept ($\mu s$)}}
\startdata
4U~1608$-$52 & 89  & $-4.9~\pm$ 0.6 & 37.3 $\pm$ 2.6\\ 
4U~1636$-$53 & 742  & $-3.2~\pm$ 0.3 & 24.3 $\pm$ 1.3\\ 
4U~0614$+$09 & 114  & $-8.1~\pm$ 1.0 & 54.4 $\pm$ 4.9\\  
4U~1728$-$34 & 174  & $-2.2~\pm$ 0.4 & 20.3 $\pm$ 1.4\\ 
4U~1702$-$43 & 112 & $-6.4~\pm$ 0.7 & 52.9 $\pm$ 3.2\\  
4U~1820$-$30 & 147 & $-1.1~\pm$ 0.4 & 8.9 $\pm$ 1.7\\ 
Aql~X-1 & 92 &  $-5.6 \pm$ 0.7 & 42.5 $\pm$ 3.3\\   
4U~1735$-$44 & 110  & $-0.9~\pm$ 0.5 & 8.1 $\pm$ 1.8\\ 
XTEJ1739-285 & 22 & $-0.5~\pm$ 1.5 & 4.0 $\pm$ 5.9\\ 
EXO1745-248 & 4 & $-1.5~\pm$ 1.1 & 11.6 $\pm$ 4.1\\ 
4U~1705$-$44 & 2 & 6.5 $\pm$ 3.9 & $-54.5~\pm$ 17.9\\ 
SAXJ1748.9-2021 & 4  & $-0.4~\pm$ 3.6 & $-2.4~\pm$ 12.4\\ 
IGRJ17191-2821 & 18  & $-0.08~\pm$ 1.41 & 2.7 $\pm$ 5.2\\ 
4U~1915$-$05 & 3 & $-4.0~\pm$ 3.7 & 35.4 $\pm$ 16.7 
\enddata
\label{tab:lagen_lower}
\end{deluxetable*}
\begin{deluxetable*}{cccc}
\tablewidth{0pt}
\tablecolumns{4}
\tabletypesize{\small}
\tablecaption{Upper kHz QPO Lag/Energy Data.}
\tablehead{
\colhead{Object ID} & \colhead{Number of Significant QPOs} & \colhead{Best Fit Slope ($\mu s/keV$)} & \colhead{Best Fit Intercept ($\mu s$)}}
\startdata
4U~1608$-$52 & 8 & $-2.3~\pm$ 1.2 & 15.0 $\pm$ 5.4\\ 
4U~1636$-$53 & 104 & 0.63 $\pm$ 0.57 & $-5.0~\pm$ 2.0\\ 
4U~0614$+$09 & 114 & $-0.5~\pm$ 1.1 & 2.3 $\pm$ 5.5\\  
4U~1728$-$34 & 130 & 0.8 $\pm$ 0.3 & $-7.2~\pm$ 1.3\\ 
4U~1702$-$43 & 112 & $-5.7~\pm$ 2.3 & 44.8 $\pm$ 8.7\\
4U~1820$-$30 & 12 & $-1.1~\pm$ 1.7 & 10.1 $\pm$ 7.0
\enddata
\label{tab:lagen_upper}	
\end{deluxetable*}

\section{Discussion} \label{discussion}
We have performed a large-scale systematic study of the spectral-timing products of kHz QPOs for neutron star LMXBs using data in the entire {\it RXTE/PCA} archive.  We examined the coherence, covariance, and lags in the lower kHz QPOs for 14 objects and in the upper kHz QPOs of 6 objects.  This is a significant increase compared to all previous studies where lags in the lower kHz QPOs have been studied in four objects \citep{Barret_13,deAvellar_13,Peille_15,deAvellar_16,Troyer_17}, and in the upper kHz QPOs for 3 objects.  We also find similar results between the various sources where we were able to compute spectral-timing products.  We summarize the main observational results:
\begin{itemize}
\item Both the lower kHz QPOs and upper kHz QPOs are highly coherent over most of the energy range studied (3 -- 20 keV).
\item The fractional covariance of all kHz QPOs generally increases with energy and levels off around 15 keV.
\item Three objects show a small decrease in lags with frequency between the 3 -- 8 keV and 8 -- 20 keV band for the lower kHz QPO, but, otherwise no statistically significant trend is found.
\item The lower kHz QPO lags generally show a decrease with energy, with the highest energy photons arriving first, while the upper kHz QPO lags are generally consistent with zero lag.
\end{itemize}
We discuss the main findings in additional detail in the following sections.

\subsection{Coherence}
While generally consistent with unity, our results show the intrinsic coherence is not well-constrained in objects with poor statistics across all energy bands.  Generally, the extreme low and high energy bins of all objects lack statistics and thus also show poorly constrained intrinsic coherence.  

The lower kHz QPOs have more and generally higher quality data, and produce more well-constrained coherence.  The confidence limits of the intrinsic coherence are difficult to quantify \citep[see e.g.,][]{Nowak_96,Vaughan_Nowak_97,Uttley_14}.  The drop in intrinsic coherence in the highest energy bin of a few sources and above $\sim$ 15 keV in 4U 0614$+$09 is likely due to the loss of QPO signal at these low S/N energies.  

\subsection{Covariance}
The covariance statistic is dependent on the coherence, and thus we expect the covariance to drop when the coherence drops.  In the regime of unity coherence, the covariance spectrum becomes equivalent to the rms spectrum with higher S/N and we generally find good agreement between the covariance and rms spectra.

The fractional covariance of all kHz QPOs generally follow a similar trend, rising with increased energy and then leveling off around 15 keV and in a few cases dropping in the highest energy bin.  An increase in fractional rms/covariance with energy is indicating that the fraction of kHz QPO signal that is variable is increasing with energy.   This is consistent with the behavior of a Comptonized emission source. \citet{Gilf_03} showed that the rms spectrum of kHz QPOs was consistent with Comptonized emission.  This idea is further supported by both \citet{Peille_15} and \citet{Troyer_17} who show the response-folded covariance spectra of kHz QPOs are consistent with Comptonized emission and do not require a disk component.  Moreover, the first application of phase-resolved spectroscopy to the lower kHz QPOs in 4U 1608$-$52 also shows that it is the Comptonized component that is modulating \citep{stevens18}.

The covariance is similar between the lower and upper kHz QPOs.  This result is consistent with \citet{Gilf_03} where they show that by subtracting a putative modeled accretion disk from the average spectrum, the result looks like the response-folded rms spectrum for all the QPOs (lower, upper, and the 45 Hz hectohertz QPO) they considered.  The fractional rms/covariance spectra are also very similar for different objects, suggesting the production mechanisms of kHz QPOs across different sources may be the same.

Our results also show the expected deviation between the covariance and rms spectra when the coherence departs from unity.  In some cases, we note drops in both rms and covariance near 20 keV.  The drop in rms and covariance in the highest energy bin of a few sources and above $\sim$ 15 keV in 4U 0614$+$09 is also likely due to the loss of QPO signal at these low S/N energies. The behavior of the covariance is similar in 4U 1728$-$34 to that described in \citet{Peille_15} using a similar aggregate data set.  \citet{Mukher_12} note a drop in fractional rms above $\sim$10 keV in 4U 1728$-$34 using a much smaller data set.

\subsection{Lags}
We analyzed the lags between the 3 -- 8 keV and 8 -- 20 keV lightcurves, determining both the average lag and looking for any frequency-dependence. Three objects (4U~1608$-$52, 4U~1820$-$30 and Aql~X$-$1) all show a decreasing trend of lags with increasing frequency for the lower kHz QPOs at $>3\sigma$ significance.  This result is generally consistent with and well-discussed in previous work \citep[see e.g.,][]{deAvellar_13}.  All other objects (for both lower and upper kHz QPOs) have no statistical significant frequency dependence, though we note that the underlying frequency-dependence of the lags may be more complex than linear.

The broad-energy lags for the lower kHz QPOs show correlated variations in the high energy band occur on average before those in the low energy band.  The average lags are on the order of a few 10s of microseconds which is consistent with the putative light-travel timescale of the inner accretion flow of a neutron star.  The average lags of the upper kHz QPOs are generally consistent with zero, and generally have fewer and lower quality data than the lower kHz QPOs.   

We also study the lags as a function of energy, averaged over all frequencies.  When comparing the energy-dependent lags between the lower and upper kHz QPOs, we notice different trends.  The lower kHz lags generally decrease monotonically as energy increases.  Those objects that do not show a strong trend typically have far fewer lower kHz QPOs detected.  The exceptions are 4U~1735$-$44 and 4U~1820$-$30, which show little energy dependence despite relatively large numbers of detections as seen in Table~\ref{tab:lagen_lower}.  Additionally, these objects have the smallest average broad energy lags of all objects with well-constrained, near unity coherence.

The upper kHz QPO lags are not as well constrained and are generally consistent with zero lag, except in the cases of 4U~1608$-$52 and 4U~1702$-$43 where the upper kHz QPOs are weakly detected.  The objects with the best upper kHz QPO photon statistics, 4U-1636$-$53 and 4U~1728$-$34, show slightly increasing lags as a function of energy.  

The energy-dependence of the lags has the potential of constraining the geometry and physics of the emission region of kHz QPOs.  For example,  in the case of reverberation, the parts of the energy spectrum that are dominated by the reflected emission should show the largest lags behind the direct emission component.  The energy-dependent lags in 4U~1608$-$52 were initially suggested to be due to reverberation by \citet{Barret_13}.   However, models of the expected reverberation lags in 4U~1608$-$52 found the energy-dependent lags diverged from the modeled lags above $\simeq$ 8 keV \citep{Cackett_16}, eliminating reverberation as the sole source of emission for the lower kHz QPO.  Moreover,  a response-folded covariance spectra, produced for lower kHz QPOs do not show an accretion disk component that would be expected from reverberation \citep{Peille_15,Troyer_17}.  Both these studies found the seed photon temperature of the covariance spectra,  which can be thought of as the spectrum of the kHz QPO, was systematically higher than the time-averaged spectra for 4U 1728$-$34 and Aql~X-1 respectively.  This is also consistent with the phase-resolved spectroscopy of 4U~1608$-$52 which shows phase lags between variations in the power-law index, high-energy cut-off temperature and seed blackbody temperature \citep{stevens18}.

\citet{Lee_Miller_98}, \citet{Lee_01}, and more recently \citet{Kumar_Misra_14,Kumar_Misra_16} have applied Comptonization models to the lag/energy and rms spectra of kHz QPOs.  Most of these studies use the rms spectrum of 4U~1608$-$52 from \citet{Berger_96} and the lag/energy spectrum from \citet{Vaughan_98}.  \citet{Kumar_Misra_14,Kumar_Misra_16} show that the lag/energy spectra of the lower kHz QPO can be recovered by a Comptonization model in which a fraction of the Comptonized emission impinges back on the seed photon source.  This includes time lags where high energy fluctuations lead lower energy fluctuations.  

\section{Conclusion} \label{conclusion}
We have analyzed all {\it RXTE/PCA} data for the objects listed in Table~\ref{tab:source_list}.  We present a large-scale, systematic comparison of spectral-timing characteristics of kHz QPOs across many sources.  We have shown that the average spectral-timing properties for all sources in the {\it RXTE/PCA} archive, which remains the best data set currently available for this kind of study, are very similar.  This supports the idea that the production mechanism of kHz QPOs is the same for all sources.  Additionally, we show that both the broad-band lags and the lag/energy spectra for the lower and upper kHz QPOs are markedly different, which supports the idea that the emission mechanisms of the two types of kHz QPOs are different.

We computed the intrinsic coherence, rms/covariance, frequency-dependent and energy-dependent lags for the lower kHz QPO in 14 objects, with 6 of those yielding results for the upper kHz QPO.  We have demonstrated only broad {\it trends} in our results.  Both high timing and higher energy resolution are required to model spectral-timing results in a robust fashion.

This work was only possible because of the existence of {\it RXTE/PCA}, which was uniquely capable of providing data necessary for spectral-timing work of kHz QPOs.   While the energy resolution coupled with variations in the detector's response limits the energy resolution of analyses, we now understand the scope of the requirements needed to fully exploit this type of analysis.  The current {\it NICER} mission can extend such analyses to lower energies than previously probed \citep{bult18}.  Future proposed missions such as {\it STROBE-X} \citep{wilsonhodge_17} and {\it eXTP} \citep{EXTP}, equipped with such capabilities will allow the full exploration of these analytic techniques.

\acknowledgements
We thank an anonymous referee for helpful comments that improved the clarity of the paper.  JST and EMC gratefully acknowledge support from the National Science Foundation through CAREER award number AST-1351222. 
\bibliographystyle{apj}
\bibliography{apj-jour,LMXB}
\end{document}